\documentstyle[12pt]{article}
\begin{document}

\title {$\sin^2\theta_W(M_Z)$ ~in
        the MSSM and Unification of Couplings
\thanks{Supported in part by the Polish Committee for Scientific
        Research.}}

\author{Piotr H. Chankowski,
Zbigniew P\l uciennik\\
and\\
Stefan Pokorski\\
\\
\\
\\
Institute of Theoretical Physics, Warsaw University\\
ul. Hoza 69, 00--681 Warsaw, Poland.
}

\maketitle

\vspace{-12cm}
\begin{flushright}
{\bf IFT--94/19}
\end{flushright}
\vspace{12cm}

\begin{abstract}
We calculate ~$\sin^2\theta(M_Z)$ ~in the MSSM in terms of ~
$\alpha_{EM}$, ~$G_F$, ~$m_t$, ~$\tan\beta$ ~and SUSY mass
parameters with the same accuracy as the present calculations
of ~$\sin^2\theta(M_Z)$ ~in the SM. We compare the results with
the standard leading logarithmic approximation used for SUSY
threshold corrections and find important differences in the case
of light sparticles. We give approximate formulae connecting
coupling constants in the SM and in the MSSM and comment on
process dependence of such formulae. The obtained values
of the MSSM couplings ~$\alpha_i(M_Z)$ ~are used to investigate gauge
coupling unification in the minimal SUSY ~$SU(5)$ ~model.
Our non-logarithmic corrections lower the predicted value of
the Higgs triplet mass. The interplay between the supersymmetric
and GUT thresholds in achieving unification for the coupling
constants in the range of the experimentally acceptable
values is quantified.
\end{abstract}

\newpage
{\bf 1. INTRODUCTION.}

\vskip 0.3cm
The succesful gauge coupling unification in the Minimal Supersymmetric
Standard Model (MSSM) \cite{UN1}
has spurred recently a lot of interest in the SUSY
GUT scenario.
The standard approach to the coupling unification is
based on the fact that physics at a given scale can be described in
terms of an effective renormalizable theory, with the effects of the
larger scales decoupled. For SUSY GUTs, one naturally considers the three
energy regimes described by the Standard Model (SM), the MSSM and the full
GUT. The first two models are effective renormalizable theories which
approximate the full theory up to higher dimension operators suppressed
by the scales of new physics.

The standard approach to the gauge coupling unification consists of the
following steps:

\noindent {\sl i}) Measurements at  energy ~$E\leq M_Z$ ~are used
         to extract the
 couplings\footnote{We call ~$\alpha_y$ ~the coupling related to
                      the ordinary hypercharge reserving the symbol~
                      $\alpha_1$ ~for the rescaled coupling appropriate
                      for discussion of unification; ~
                      $\alpha_1 = (5/3)\alpha_y$.} ~
        $\alpha_i(M_Z)$ ~$i=y,2,3$ ~using the Dimensional Reduction
  regularization (DR)
        \cite{DRED} and the Modified Minimal
        Subtraction Scheme ~($\overline{MS}$ \cite{MS_BAR})
        in the framework of the SM. In addition
        to ~$\alpha_3(M_Z)$ ~and~ $\alpha_{EM}(M_Z)$ ~it is convenient
        to use ~$\sin^2\theta_W(M_Z)$.
        It can be calculated e.g. in terms
        of ~$G_F=1.16639(2)\times10^{-5}$ GeV$^{-2}$, ~
        $\alpha_{EM}^{OS}=1/137.0359895(61)$, ~$M_Z=91.1888(44)$ GeV ~
        \cite{GLA}, ~$m_t$ ~and ~$M_{\phi^0}$~
        (the top quark and SM Higgs boson masses respectively).
        The first three parameters are known with very high accuracy
        and the latter two enter through radiative corrections. In
        practice, one takes ~$\alpha_{EM}(M_Z)$, ~$\sin^2\theta_W(M_Z)$ ~and~
        $\alpha_3(M_Z)$ ~determined in the ordinary Dimensional
        Regularization (DIMR) and $\overline{MS}$
        and subsequently converts them to the
        DR regularization using the prescription \cite{CONV}
\begin{eqnarray}
        {1\over\alpha_i(M_Z)} = {1\over\alpha_i^{DIMR}(M_Z)} - {C_i\over12\pi}
\label{eqn:conv}
\end{eqnarray}
        with ~$C_y=0, ~C_2=2, ~C_3=3$. The Weinberg angle can also be
        extracted (or crosschecked) from other observables: e.g. ~$G_F$ ~
        can be replaced by ~$\sin^2\theta_W^{eff}$ ~measured
        at LEP and SLD \cite{GLA}. However, at present the most precise
        determination of the couplings in the SM follows from the measured
        values of ~$G_F$, ~$M_Z$ ~and ~$\alpha_{EM}^{OS}$.

\noindent {\sl ii}) The coupling constants are evolved to higher scales
        by means of the 2--loop RGE of the SM, up to some scale ~$M_{SUSY}$,
        and then using the RGE of the MSSM. The threshold corrections
        to the running of ~$\alpha_i(Q)$ ~due to the splitting in
        sparticle masses are included at the 1-loop level and result in the
        contribution \cite{RR,LP,CARENA}
\begin{eqnarray}
{1\over\alpha^{MSSM}_i(M_{SUSY})}
- {1\over\alpha^{SM}_i(M_{SUSY})}\equiv4\pi\Delta_i(M_{SUSY})
                           = \sum_k \Delta b_{ik} \log{m_k\over M_{SUSY}}
\label{eqn:thrs}
\end{eqnarray}
        where the coefficients ~$\Delta b_{ik}$ ~are the contrbutions
        to the ~$\beta_i$ ~function of the coupling ~$\alpha_i$ ~from
        the sparticle ~$k$. ~
        It has to be stressed, that the effects of the VEV contribution to
        sparticle masses as well as the effects of the {\sl left-right}
        mixing between sfermions cannot be treated consistently in this
        way and must be neglected. Thus the masses ~$m_k$ ~appearing in
        the formula (\ref{eqn:thrs}) are the lagrangian ~$SU(2)\times U(1)$
        invariant mass parameters and not the physical masses.

The one loop formula for ~$\alpha^{MSSM}_i(Q)$ ~at some scale ~$Q > max(m_k)$:
\begin{eqnarray}
{1\over\alpha_i^{MSSM}(Q)} = {1\over\alpha^{SM}_i(M_Z)}
 &+& 2b^{SM}_i \log{M_{SUSY}\over M_Z}\nonumber\\
 &+& 4\pi\Delta_i(M_{SUSY}) + 2b^{MSSM}_i \log{Q\over M_{SUSY}}
\end{eqnarray}
can be rewritten as
\begin{eqnarray}
{1\over\alpha_i^{MSSM}(Q)} = {1\over\alpha^{MSSM}_i(M_Z)} +
2b^{MSSM}_i \log{Q\over M_Z}
\end{eqnarray}
with
\begin{eqnarray}
{1\over\alpha^{MSSM}_i(M_Z)} = {1\over\alpha^{SM}_i(M_Z)}
+ 4\pi\Delta_i(M_Z)
\label{eqn:runmatch}
\end{eqnarray}
having the obvious interpretation of the couplings ~
$\alpha_i(M_Z)$ ~extracted at the scale ~$M_Z$ ~directly
in the MSSM. This is possible, because the threshold corrections are
given in eq.(2) in the form in which the 1-loop logarithms are resummed.
Therefore the dependence on ~$M_{SUSY}$ ~in eq.(3) drops out and one
can take  ~$M_{SUSY}=M_Z$. One should stress that this is true only in
the leading logarithmic approximation for the threshold corrections.
In general the results would depend on the scale $M_{SUSY}$~ and this
dependence is always a next order correction.

\noindent {\sl iii}) The MSSM equations are used for the running up to some
        scale ~$M_{GUT}$ ~where the unification is achieved (or not) without
        or with the GUT scale threshold effects which are taken into
      account also in the leading logarithmic approximation.
      In the latter case the
        final result depends on the specific GUT. In addition it depends on
        our criterion of an ``acceptable'' GUT spectrum.

The standard approach is justified in the
presence of well separate scales \newline
($M_Z$ $\ll M_{SUSY}\ll M_{GUT}$). ~
However, it often happens that the scales
of ``new physics'' are not large enough to neglect its non--logarithmic
contributions. This may well be the case with the Standard Model and its
supersymmetric extension: it is quite plausible that the superpartner masses
vary from ~${\cal O}(M_Z)$ ~(some of them can even be lighter than ~$Z^0$~
boson) to, say, ~${\cal O}$(1 TeV). ~In this case the (renormalizable) SM
is not a consistent approximation at the scale ~$M_Z$ ~to the full MSSM
and higher dimension (non-renormalizable)  operators cannot be neglected
in step {\sl i}), i.e. in the procedure of extracting the SM couplings from
the data.

An equivalent,  more straightforward and more convenient approach is to work
at the scale ~$M_Z$ ~directly with the MSSM. This point has been
emphasized by Faraggi and Grinstein \cite{FAGRI}.
This means that the couplings
at ~$M_Z$ ~extracted from experimental data are the MSSM (and not the SM)
couplings. All SUSY threshold effects are already included at this step and
the RG running is based on 2-loop MSSM equations from ~$M_Z$ ~up to the
GUT scale.
{}From this point of view formula (\ref{eqn:runmatch}) is a formula for ~
$\alpha_i^{MSSM}(M_Z)$ ~calculated e.g. in terms of~
$G_F$, ~$\alpha_{EM}^{OS}$, ~$M_Z$, ~$m_t$ ~and other SUSY
parameters with only 1-loop logarithmic SUSY contributions retained, and
resummed to all orders. We call it Leading Logarithmic Threshold (LLT)
Approximation.

The main part of this paper is the calculation in Sec. 2
of the ~$\sin^2\theta_W(M_Z)$ ~(and also,
as a byproduct, of ~$M_W$)~ in terms of ~$\alpha_{EM}^{OS}$,~
$G_F$, ~$M_Z$, ~$m_t$, ~$\tan\beta$ ~and other SUSY parameters in the ~
$\overline{MS}$ \cite{MS_BAR} Scheme directly in the MSSM with the same
accuracy as the present calculations of ~$\sin^2\theta_W(M_Z)$ ~in
the SM. In Sec. 3 we compare the results with the
standard LLT approximation and find important differences in the case of light
sparticles. Also in Sec. 3 we comment on the Farragi-Grinstein (FG)
\cite{FAGRI} approach and its ambiguity related to the process dependence.
In an approximation analogous to
theirs we give the formulae connecting the coupling constants in the SM
and in the MSSM, if calculated in terms
of ~$G_F$, ~$M_Z$ ~and $\alpha_{EM}^{OS}$. ~In Sec.4 we use the obtained
values of the MSSM couplings ~$\alpha_i(M_Z)$ ~to investigate the gauge
coupling unification in the minimal SUSY ~$SU(5)$ ~model.
In particular, we find
that our non-logarithmic corrections lower the predicted values
of the Higgs triplet mass. Also, we study in some detail the
interplay between the supersymmetric and GUT thresholds
in achieving unification for the coupling constants in the range
of the experimentally acceptable values.

\vskip 0.5cm
{\bf 2. CALCULATION OF  ~$\sin^2\theta_W(M_Z)$ ~IN THE MSSM.}
\vskip 0.3cm

The calculation of ~$\sin^2\theta_W(M_Z)$ ~proceeds
through the calculation
of the Fermi constant measured in the ~$\mu\rightarrow e\bar\nu_e\nu_{\mu}$~
decay.
We follow closely the analogous calculation in the SM whose details
are analyzed in  \cite{DFS}. The equation
for ~$\hat s^2\equiv\sin^2\theta_W(M_Z)$ ~
reads (for the running coupling constants in the MSSM
taken at the scale ~$Q=M_Z$ ~we use the abbreviations: ~
$\hat\alpha\equiv\alpha_{EM}(M_Z)$ ~and ~
$\hat\alpha_i\equiv\alpha_i(M_Z)$ ~for i=y,2,3;
analogous couplings in the SM will be distinguished always
by the superscript ``SM''.):
\begin{eqnarray}
\hat s^2 = {\pi\hat\alpha\over\sqrt2 G_F M^2_W}
{1\over 1-\Delta\hat r_W(\hat\alpha,M_Z,M_W,m_t,...)}
\label{eqn:s2def}
\end{eqnarray}
where
\begin{eqnarray}
\Delta\hat r_W = -{\hat\Pi_{WW}(0)\over M^2_W}
               + {\hat\Pi_{WW}(M^2_W)\over M^2_W} + \hat\delta_{VB}
\label{eqn:drdef}
\end{eqnarray}
and ~$M_W$ ~($M_Z$) ~is the physical mass of the ~$W^{\pm}$ ~($Z^0$) ~boson.
The ~$\hat\Pi_{V_1V_2}(q^2)$'s ~are the vector boson self energies
calculated in DR and renormalized by the Modified Minimal Subtraction ~
$\overline{MS}$; ~
the expressions for ~$WW$, ~$ZZ$, ~$\gamma\gamma$ ~and ~$Z\gamma$ ~
self energies in ~DR ~can be found in \cite{MY_H}) and ~
$\hat\delta_{VB}=\hat\delta^{SM}_{VB}+\hat\delta^{SUSY}_{VB}$ ~is the
contribution of ``non-oblique'' vertex, box and external fermion
wave function renormalization factors. In the ~$\overline{MS}$ ~scheme,
the SM contribution
$\hat\delta^{SM}_{VB}$  is given by \cite{DFS}:
\begin{eqnarray}
\hat\delta^{SM}_{VB}= {\hat\alpha\over4\pi\hat s^2}
\left[6+
{7-5s^2_W + \hat s^2\left(3c^2_W/\hat c^2 -10\right)\over2s^2_W}\log c^2_W
\right]
\label{eqn:box}
\end{eqnarray}
with ~$c^2_W = 1 - s^2_W$ ~used as an
abbreviation for ~$M^2_W/M^2_Z$. ~
It has the same form in the ~DIMR ~and ~DR ~schemes.
The ~$\hat\delta^{SUSY}_{VB}$ ~has exactly the same form
as in the on-shell calculation (with the obvious
replacement of ~$s_W$, ~$c_W$ and ~$\alpha_{EM}^{OS}$ ~by~
$\hat s$, ~$\hat c$ ~and ~$\hat\alpha$) ~
and can be taken over from the Appendix A of ref. \cite{MY_DR}: the
counterterms to the vertices and neutrino self energies have in the~
$\overline{MS}$  ~calculation the interpretation of the electron and muon
wave function renormalization factors.

One way to organize the calculation with ~$G_F$, ~$M_Z$ ~and ~
$\alpha_{EM}^{OS}$ ~taken as the input parameters is  \cite{DFS}
to express the physical mass of the ~$W^{\pm}$ ~boson in terms of ~$M_Z$:
\begin{eqnarray}
M^2_W = \hat c^2\hat\rho M^2_Z
\label{eqn:mwdef}
\end{eqnarray}
with
\begin{eqnarray}
{1\over\hat\rho} =
{{\displaystyle 1- {\hat\Pi_{WW}(M_W^2)\over M^2_W}}\over
 {\displaystyle 1- {\hat\Pi_{ZZ}(M_Z^2)\over M^2_Z} -
{\delta\hat\Pi_{ZZ}(M_Z^2)\over M^2_Z}}}
+ \delta\hat\rho
\label{eqn:rodef}
\end{eqnarray}
where ~$\delta\hat\Pi_{ZZ}(M_Z^2)$ ~is the contribution of the ~$Z-\gamma$ ~
mixing to the relation between the physical and bare ~$Z^0$ ~boson masses:
\begin{eqnarray}
\delta\hat\Pi_{ZZ}(M_Z^2)=
{\left(\hat\Pi_{Z\gamma}(M_Z^2)\right)^2\over
M^2_Z - \hat\Pi_{\gamma\gamma}(M_Z^2)}
\end{eqnarray}
and ~$\delta\hat\rho = \delta\hat\rho^{QCD} + \delta\hat\rho^{HIGGS}$ ~
stands for the leading irreducible 2-loop corrections to ~$\hat\rho$ ~of
order ~${\cal O}(\alpha_3 G_F m^2_t)$ ~(QCD) ~and ~
${\cal O}(G_F^2 m^4_t)$ ~(HIGGS).  ~$\delta\hat\rho^{QCD}$ ~in the SM
and in the limit of a heavy top quark reads \cite{QCD}
\begin{eqnarray}
\delta\hat\rho^{QCD} =
{2\alpha_3(m_t)\over3\pi}\left(1+{\pi^2\over3}\right)
N_c\left({G_Fm^2_t\over 8\sqrt2\pi^2}\right)
\end{eqnarray}
where ~$m_t$~ is the top quark pole mass.
This expression is expected to account for the dominant part of QCD
corrections also in the MSSM since corrections coming from gluons attached to
squark loops, which themselves contribute a small piece of ~$\hat\rho$, ~
should be very small. ~In the SM, ~$\delta\hat\rho^{HIGGS}$ ~
was calculated in ref \cite{BIJ} ~in the limit ~
$M_{\phi^0}=0$. Recently this calculation
has been extended to ~$M_{\phi^0}\neq0$ ~by Barbieri et al. \cite{BARB}.
In the absence of explicit calculation of  ~
$\delta\hat\rho^{HIGGS}$ ~in the MSSM we use the following interpolating
formula:
\begin{eqnarray}
\delta\hat\rho^{HIGGS} = - N_c\left({G_Fm^2_t\over 8\sqrt2\pi^2}\right)^2
\left[\left({\cos\alpha\over\sin\beta}\right)^2
\rho^{(2)}\left({M_{h^0}\over m_t}\right)
+\left({\sin\alpha\over\sin\beta}\right)^2
\rho^{(2)}\left({M_{H^0}\over m_t}\right)\right]\nonumber\\
\label{eqn:dro_higgs}
\end{eqnarray}
where ~$\rho^{(2)}(x)$ ~is given in \cite{BARB} (see also \cite{HOLL}),~
$M_{h^0}$ ~and ~$M_{H^0}$ ~are  masses of the lighter and heavier MSSM
scalar Higgs bosons, ~$\tan\beta\equiv v_2/v_1$ ~and ~$\alpha$ ~is
the mixing angle between scalar Higgs bosons.

Finally, the running coupling constant ~$\hat\alpha\equiv\alpha(M_Z)$ ~is
obtained from the formula \cite{HOLL}:
\begin{eqnarray}
\hat\alpha = {\alpha\over 1-\Delta\hat\alpha}
\label{eqn:alhat}
\end{eqnarray}
with ~$\alpha\equiv\alpha^{OS}_{EM}$ ~and
\begin{eqnarray}
\Delta\hat\alpha = 0.0684\pm0.0009
+{7\alpha\over4\pi}\log{M_W\over M_Z}
-{8\alpha\over9\pi}\log{m_t\over M_Z}
-4\pi\alpha~\Delta_e
\label{eqn:dalfa}
\end{eqnarray}
where
\begin{eqnarray}
8\pi^2\Delta_e &=& \sum_{i=1}^2 {4\over3}\log{m_{C_i}\over M_Z}
         + {1\over3}\log{M_{H^{\pm}}\over M_Z}\nonumber\\
         &+& \sum_{i=1}^6 {1\over3}\log{M_{E_i}\over M_Z}
         +   \sum_{i=1}^6 {1\over9}\log{M_{D_i}\over M_Z}
         +   \sum_{i=1}^6 {4\over9}\log{M_{U_i}\over M_Z}
\label{eqn:deltae}
\end{eqnarray}
contains logarithms of the physical masses of all
charged supersymmetric particles.
Notice, that as compared to the DIMR (see e.g. \cite{HOLL}), the quantity~
$\Delta\hat\alpha$ ~in DR does not contain any constant term
coming from integrating out ~$W^{\pm}$ ~and ~$Z^0$ ~bosons
\footnote{This is the content of the theorem \cite{CONV} stating that in
the ~DR ~scheme there are no threshold corrections when gauge bosons are
integrated out.
}

The term ~$0.0684\pm0.0009$ ~accounts for the contribution of light
fermions and contains a nonperturbative part. It is extracted from the
experimental data on ~$e^+e^-\rightarrow\gamma^*\rightarrow hadrons$
\cite{JEG}. The error 0.0009 is the most important theoretical
uncertainity in the determination of ~$\hat s^2$ from
eq. (\ref{eqn:s2def}).

The fine structure constant ~$\alpha_{EM}^{OS}$ ~is obtained from
the Thomson scattering
in pure QED, which is the effective theory  at the scale
of the electron mass. At this scale, all higher dimensional operators
which remain after integrating out all heavier particles are totally
negligible and so  are the corrections to  eq.(\ref{eqn:alhat}).

 Eq.(\ref{eqn:s2def})  is solved iteratively by calculating
at each step of iteration the physical ~$W$ ~boson mass from
eq.(\ref{eqn:mwdef})  and ~$\hat\alpha$ ~from eq.(\ref{eqn:alhat}).

In the limit of very heavy sparticles  SUSY corrections to
self energies of the vector
bosons are dominated by large logarithms of the soft SUSY breaking
masses. In this limit our approach should give the same results as given
in eq.(\ref{eqn:runmatch}) in the LLT approximation, where the logarithms
read explicitly:
\begin{eqnarray}
8\pi^2\Delta_y &=& {1\over3}\sum_{i=1}^3 \log{m_{E_i}\over M_Z}
                +  {1\over6}\sum_{i=1}^3 \log{m_{L_i}\over M_Z}\nonumber\\
               &+& {1\over9}\sum_{i=1}^3 \log{m_{D_i}\over M_Z}
                +  {4\over9}\sum_{i=1}^3 \log{m_{U_i}\over M_Z}
                +  {1\over18}\sum_{i=1}^3 \log{m_{Q_i}\over M_Z}\nonumber\\
               &+& {2\over3}\log{|\mu|\over M_Z}
                +  {1\over6}\log{M_{A^0}\over M_Z}
\label{eqn:deltay}
\end{eqnarray}
\begin{eqnarray}
8\pi^2\Delta_2 &=& {1\over6}\sum_{i=1}^3 \log{m_{L_i}\over M_Z}
                +  {1\over2}\sum_{i=1}^3 \log{m_{Q_i}\over M_Z}\nonumber\\
               &+& {2\over3}\log{|\mu|\over M_Z}
                +  {1\over6}\log{M_{A^0}\over M_Z}
                +  {4\over3}\log{M_{g2}\over M_Z}
\label{eqn:delta2}
\end{eqnarray}
(~$m_{Q_i}, m_{L_i}$, ... are the soft SUSY breaking mass parameters).
{}From eq.(\ref{eqn:runmatch})  one easily gets:
\begin{eqnarray}
{1\over\hat\alpha} = {1\over\hat\alpha^{SM}} + 4\pi (\Delta_y + \Delta_2)
\label{eqn:alsum}
\end{eqnarray}
\begin{eqnarray}
\hat s^2 = \hat s^2_{SM} {1 + 4\pi\hat\alpha_2^{SM}\Delta_2
           \over 1 + 4\pi\hat\alpha^{SM}\left(\Delta_y + \Delta_2\right)}
\label{eqn:s2sum}
\end{eqnarray}
where quantities with the superscript ~$SM$ ~are determined by  equations
(\ref{eqn:s2def}--\ref{eqn:dalfa})
in the SM i.e. with the contributions of SUSY particles removed.
Analogous formula with ~$\alpha_2^{SM}\Delta_2$ ~replaced by~
$\alpha_y^{SM}\Delta_y$ ~holds for ~$\hat c^2$. ~Since
in the limit of heavy sparticles the physical masses are dominated
by large soft SUSY breaking masses we have ~
$\Delta_e = \Delta_y + \Delta_2 + {\cal O}({M^2_W\over M^2_{SUSY}})$ ~and
it is easy to see that indeed ~$\hat\alpha$ ~calculated from
eq.(\ref{eqn:alhat})  approaches ~$\hat\alpha$ ~determined from
eq.(\ref{eqn:alsum}). Next, it is important to notice that in the
limit considered, by using the asymptotic form of the SUSY contributions to
self energies
\footnote{Subtraction of ~$\hat\Pi(0)$ ~for ~$W^{\pm}$ ~and ~$Z^0$ ~self
energies is required because of the chargino/neutralino contrubution which
unlike scalars do not decouple in that limit from renormalized self--energies
at ~$q^2=0$; ~however, ~
$\hat\Pi_{WW}(0)/M^2_W - \hat\Pi_{ZZ}(0)/M^2_Z$ ~vanishes in the limit
of heavy charginos and neutralinos as it should.}:
\begin{eqnarray}
\hat\Pi_{WW}(q^2)-\hat\Pi_{WW}(0)
&\sim&4\pi{\hat\alpha\over\hat s^2} ~q^2~\Delta_2,\nonumber\\
\hat\Pi_{ZZ}(q^2)-\hat\Pi_{ZZ}(0)
&\sim&4\pi{\hat\alpha\over\hat s^2\hat c^2} ~q^2~
(\hat c^4\Delta_2 + \hat s^4\Delta_y),\nonumber\\
\hat\Pi_{Z\gamma}(q^2)&\sim&4\pi{\hat\alpha\over\hat s\hat c} ~q^2~
(\hat c^2\Delta_2 - \hat s^2\Delta_y),\\
\hat\Pi_{\gamma\gamma}(q^2)&\sim&4\pi\hat\alpha ~q^2~
(\Delta_2 + \Delta_y).\nonumber
\label{eqn:piasy}
\end{eqnarray}
eq.(\ref{eqn:s2def}) can be rewritten in the following form:
\begin{eqnarray}
\hat s^2 = \hat s^2_{SM} {1 + 4\pi\hat\alpha_2^{SM}\Delta_2
           \over 1 + 4\pi\hat\alpha^{SM}\left(\Delta_y + \Delta_2\right)}
           {1\over1  - \Delta\hat r_{SM} 4\pi\hat\alpha_2^{SM}\Delta_2}
\label{eqn:mod_s2def}
\end{eqnarray}
which is the same as eq.(\ref{eqn:alsum}) (up to higher order terms).
Thus, we reproduce correctly the leading logarithmic threshold
corrections in the limit of heavy sparticles.

With regard to eqs.(\ref{eqn:mwdef}) and (\ref{eqn:rodef}) we see that
they connect
with each other two observables ~$M_W$ ~and ~$M_Z$. ~Therefore,
according to the Appelquist--Carazzone decoupling  theorem,
heavy SUSY particles have no effect on this relation; one can
explicitely check this cancellation in the limit of heavy sparticles
with the help of eqs.(21) and (\ref{eqn:mod_s2def}).

Before presenting our results for ~$\sin^2\theta(M_Z)$ ~(see Sec. 3)
we comment on various consistency checks of our calculation.
The higher order correction in eq.(\ref{eqn:s2def}) coming from
neglected (non--top exchange)
2--loop diagrams can be estimated as in the SM \cite{HOLL}
to be of order ~$\delta\hat s^2 < 3\times10^{-5}$; ~
the uncertainty due to the difference between
eqs.(\ref{eqn:s2def}) and (\ref{eqn:mod_s2def})
turns out to be ~$\leq1\times10^{-4}$~
for ~$M_{SUSY}\leq 1.5$ TeV.
The ~$W^{\pm}$~ mass calculated in the MSSM in ~${\overline{MS}}$ ~
scheme (eqs.(\ref{eqn:mwdef}) and (\ref{eqn:rodef}))
and in the on-shell scheme\footnote
{In the present version, our code used in \cite{MY_DR} has been
          improved to incorporate ~$\delta\rho^{HIGGS}$ ~as in
          eq.(\ref{eqn:dro_higgs}) and the
          leading top dependent corrections to the Higgs boson masses
          using the effective potential approach \cite{EZ} as described in
          \cite{JA_STOP}.}
\cite{MY_DR} agree to better than ~
${\cal O}(10)$ MeV ~for a  wide range of sparticle masses. The ~
$W^{\pm}$ ~mass calculated in the MSSM (by using
eqs.(\ref{eqn:mwdef}) and (\ref{eqn:rodef})) with sparticle
masses  greater than~ 1.5 TeV ~agrees  to better than ~10 MeV ~with the ~
$W^{\pm}$ ~mass calculated in the SM  (in the ~$\overline{MS}$ ~scheme
or in the on--shell sheme) with the appropriate SM Higgs
boson mass ~$M_{\phi^0}$.

\newpage
{\bf 3. COMPARISON WITH THE LEADING LOGARITHMIC
 AND FARAGGI-GRINSTEIN  APPROXIMATIONS.}
\vskip 0.3cm

We begin the presentation of our results
with a brief summary on  ~$\hat s^2_{SM}$ ~($\sin^2\theta_W(M_Z)$ ~in
the Standard Model) calculated with our  code
in terms of  ~$G_F$, ~$M_Z$ ~
$\alpha\equiv\alpha_{EM}^{OS}$, ~
$m_t$ ~and ~$M_{\phi^0}$. A comparison
with earlier results \cite{DFS,HOLL} is a useful  check of our calculation
and remembering several numerical values provides a convenient reference
frame for comparison  with the MSSM results. In Table 1 we collect a sample
of values  for ~$\hat s^2_{SM}$ (in DIMR ) and ~$M_W$:
\vskip 0.3cm
\begin{center}
{\bf Table 1}
\vskip 0.2cm
\begin{tabular}{||l||l|l|l|l|l|l||} \hline
$m_t$        & 160    & 160    & 170    & 170    & 180    & 180 \\ \hline
$M_{\phi^0}$ & 60     & 130    & 60     & 130    &  60    & 130 \\ \hline
$s^2_{SM}$   & 0.23155& 0.23195& 0.23125& 0.23164& 0.23093& 0.23132\\ \hline
$M_W$        & 80.344 & 80.301 & 80.406 & 80.362 & 80.471 & 80.426 \\ \hline
\end{tabular}
\end{center}
\vskip 0.3cm
We reproduce the results of ref. \cite{DFS}
\footnote{Provided as in ref. \cite{DFS} we take ~$M_Z$=91.17 GeV,
{}~$\delta\rho^{QCD}=0$,~
          $\delta\rho^{HIGGS}=2\pi^2-19$, ~and adjust the hadronic
          vacuum polarization to their value.}
within $\pm 0.00008$~ accuracy in ~$\hat s^2_{SM}$~ for a wide range of ~$m_t$
and ~$M_{\phi^0}$~values.
Our input parameters (with which the Table 1 has been obtained) are: ~
$M_Z=91.1888$ GeV, ~$G_F=1.166739\times10^{-5}$ GeV$^{-2}$,~
$\alpha\equiv\alpha_{EM}^{OS} = 1/137.0369895$ ~and ~$\alpha_3(M_Z)=0.118$.
The running electromagnetic coupling ~
$\hat\alpha^{SM}\equiv\alpha_{EM}(M_Z)$  ~in the ~$\overline{MS}$~
scheme reads:\begin{eqnarray}
{1\over\hat\alpha^{SM}} = 127.87 + {8\over9\pi}\log{m_t\over M_Z}
\end{eqnarray}
and is calculated from eq.(\ref{eqn:alhat}) by taking
\begin{eqnarray}
\Delta\hat\alpha = 0.0684
-{\alpha\over6\pi}
+{7\alpha\over4\pi}\log{M_W\over M_Z}
-{8\alpha\over9\pi}\log{m_t\over M_Z}
\end{eqnarray}
appropriate for the calculation in DIMR  (the conversion to
DR can always be performed with the help of eq.(\ref{eqn:conv})).
The uncertainty in the value of ~$\hat\alpha^{SM}$~ related to the
hadronic contribution to ~$\Delta\hat\alpha$ ~
($\delta(\Delta\hat\alpha) = \pm0.0009$ ~\cite{JEG}) is a source of
uncertainities: ~$\delta\hat s^2=\pm 0.0003$ ~and ~$\delta M_W =
\pm 17$ MeV.

Similarily, our values of the physical mass ~$M_W$ ~obtained in the
Standard Model from eqs.(\ref{eqn:mwdef}) and (\ref{eqn:rodef})
agree very well
with the previous calculation \cite{DFS,HOLL} (up to a  few MeV).
As mentioned earlier, the present calculation in the ~$\overline{MS}$~
scheme also agrees with the calculation in the on shell
renormalization scheme: for ~$90 < m_t < 240$ GeV ~and ~
$60 < M_{\phi^0} < 1000$ GeV ~the difference is less than 5 MeV  and for ~
$m_t < 210$ GeV ~and ~$M_{\phi^0} < 500$ GeV ~the agreement is typically
within 2 MeV.

For later use it is very convenient to have simple interpolating formulae
for ~$\hat s^2_{SM}$ ~and ~$M_W$ ~in the SM. For the central values
of the input parameters
we find that the following formulae:
\begin{eqnarray}
\hat s^2_{SM}= 0.23166 &+& 5.4\times10^{-6} h
                        - 2.4\times10^{-8} h^2\nonumber\\
                        &-& 3.03\times10^{-5} t
                        - 8.4\times10^{-8} t^2
\label{eqn:fits}
\end{eqnarray}
\begin{eqnarray}
(M _W)_{SM} = 80.347 &-&6.4\times10^{-4} h
                       +2.5\times10^{-6} h^2\nonumber\\
                        &+&6.2\times10^{-3} t
                        +1.25\times10^{-5} t^2~,
\label{eqn:fitw}
\end{eqnarray}
where ~$h\equiv M_{\phi^0}-100$ ~and ~$t\equiv m_t - 165$ ~(both masses
in GeVs), reproduce the results for ~$\hat s^2_{SM}$ ~and ~$M_W$ ~in the SM
with the accuracy 0.00001 and 2 MeV respectively (for the range ~
$130 < m_t < 200$ GeV ~and ~$60 < M_{\phi^0} < 140$ GeV ~which is relevant
for comparison with MSSM). ~These formulae
are therefore used as input in eq.(\ref{eqn:runmatch})
in our further study whenever we refer to the values
of ~$\hat s^2_{SM}$ ~and ~$M_W$ ~ in the Standard Model.

Turning now to our results for ~$\hat s^2$ ~in the MSSM, there are two
points to be discussed first: the departure from the SM result (with the
same values of the input parameters ~$G_F$, ~$M_Z$, ~$\alpha_{EM}^{OS}$, ~
$m_t$ ~and ~$M_{h^0}$)  ~as a function of the SUSY
particle masses and comparison with the LLT approximation \cite{RR,JAP}.
Both are illustrated in Fig.1 a-d where we present the results of our
complete calculation and of the LLT approximation for a scan over the
sparticle masses in a wide range of values given in the Appendix.
In our scan we include only those values of the parameters
which are consistent with experimental constraints on the sparticle
masses, the mass of the lighter MSSM scalar
Higgs boson \cite{JA_STOP} and give~$M_W$ ~
within ~$2\sigma$~ of the measured value \cite{MW}.
In Fig.1, plots for ~$m_t=180$ GeV ~contain
smaller number of points because most of the cases with light sparticles
are eliminated by the
mentioned above experimental cuts \cite{MY_DR,JA_STOP}.
All points for ~$m_t=160$ (180) GeV ~in Fig.1 correspond to
the values ~$\hat s^2_{SM} = 0.23155 - 0.23195$ ~
($0.23093 - 0.23132$), ~with the ranges reflecting the (weak) dependence
of the ~$\hat s^2_{SM}$ ~on ~$M_{\phi^0}$ ~in the range ~$60 - 130$ GeV
(see Table 1)
\footnote{This dependence is neglected in the frequently used
fit ~$\hat s^2_{SM} = 0.2326 - 1.03\times10^{-7}(m_t^2 - 138^2)$
\cite{LP}.}
and show the dependence of the running couplings
in the MSSM on SUSY thresholds.

Comparing the complete calculation with the LLT approximation we see that
the difference in ~$\hat s^2$ ~is up to ~${\cal O}(0.002)$ ~for light
supersymmetric spectrum and therefore very relevant for the
discussion of unification of couplings.
There is also a nonnegligible dependence of ~$\hat\alpha$ ~on the
contribution from the electroweak symmetry breaking to the charged
sparticle masses  (see Figs.2-4).

The origin of the differences between the full and approximate calculations
can be understood from formulae  quoted in the Appendix. Qualitatively,
the largest deviations from the LLT approximation come from the
contribution of sfermion and, to the smaller extent, chargino/neutralino
sectors. This is further illustrated in
Figs.2-4 where we show the dependence of our results on the chosen
sparticle masses keeping other SUSY particles heavy ~(${\cal O}$(1 TeV)).

Finally we discuss the approximate method proposed by Faraggi and
Grinstein \cite{FAGRI} of translating the values of the couplings in
the SM into the values of the couplings in the MSSM, without performing
the complete calculation in the MSSM, as in Sec. 2.

In this approach  one is instructed to choose a set of observables ~
${\cal O}_i$ ~which can be calculated in the SM and MSSM  as
\begin{eqnarray}
{\cal O}_i = F_i(g^{SM}_k)
\end{eqnarray}
and
\begin{eqnarray}
{\cal O}_i = \tilde F_i(g^{MSSM}_k, e_n)
\end{eqnarray}
respectively,
where ~$e_n$ ~stand for SUSY parameters not present in the SM. Writing
next ~$g^{MSSM}_k = g^{SM}_k + \delta g_k$, ~$\tilde F_i(g^{MSSM}_k, e_n)
= F_i(g^{MSSM}_k) + \Delta F_i(g^{MSSM}_k, e_n)$
 ~and equating the RHS of the
above two equations one arrives (assuming that ~$\delta g_k$ ~are small
enough) at the set of equations
\begin{eqnarray}
\sum_{l}\delta g_l {\partial F_i(g_k^{SM})\over\partial g^{SM}_l}
+ \Delta F_i(g_k^{SM}, e_n) = 0
\end{eqnarray}
which allow for the determination of the MSSM couplings provided the SM
couplings ~$g^{SM}_k$ ~have been determined from the {\it chosen set of
observables} ~${\cal O}_i$.

This prescription can be easily applied to our set of observables: ~
$\alpha_{EM}^{OS}$, ~$G_F$ ~and ~$M_Z$ ~leading to the following approximate
formulae (which can be also derived directly from eqs.(6-10) without
any reference to the FG method):
\begin{eqnarray}
{\delta g_2^2\over g_{2,SM}^2} = {1\over \hat c^2_{SM} - \hat s^2_{SM}}
\left[4\pi\hat\alpha^{SM}\hat s^2_{SM} \Delta_e
+ \hat c^2_{SM} {\cal R}e\left({\hat\Pi_{WW}(0)\over M^2_W}
- {\hat\Pi_{ZZ}(M_Z^2)\over M^2_Z}\right)^{SUSY}\right]\nonumber
\end{eqnarray}
\begin{eqnarray}
{\delta g_y^2\over g_{y,SM}^2} = {1\over \hat s^2_{SM} - \hat c^2_{SM}}
\left[4\pi\hat\alpha^{SM}\hat c^2_{SM} \Delta_e
+ \hat s^2_{SM} {\cal R}e\left({\hat\Pi_{WW}(0)\over M^2_W}
- {\hat\Pi_{ZZ}(M_Z^2)\over M^2_Z}\right)^{SUSY}\right]\nonumber\\
\end{eqnarray}
or, equivalently
\begin{eqnarray}
{\delta\hat\alpha\over\hat\alpha^{SM}} = - 4\pi\hat\alpha^{SM}\Delta_e
\label{eqn:damy}
\end{eqnarray}
\begin{eqnarray}
{\delta\hat s^2\over\hat s^2_{SM}}=
{\hat c^2_{SM}\over\hat s^2_{SM}-\hat c^2_{SM}}
\left[4\pi\hat\alpha^{SM}\Delta_e
+ {\cal R}e\left({\hat\Pi_{WW}(0)\over M^2_W}
- {\hat\Pi_{ZZ}(M_Z^2)\over M^2_Z}\right)^{SUSY}\right]
\label{eqn:dsmy}
\end{eqnarray}
Similarily, one can easily find the formula for correction to the $W^{\pm}$~
mass ~($\delta M^2_W\equiv (M^2_W)_{MSSM} - (M^2_W)_{SM}$):
\begin{eqnarray}
{\delta M^2_W\over (M_W^2)_{SM}} =
-{\hat s^2_{SM}\over\hat c^2_{SM}}
\left({\delta\hat s^2\over\hat s^2_{SM}}\right)
+ {\cal R}e\left({\hat\Pi_{WW}(M^2_W)\over M^2_W}
- {\hat\Pi_{ZZ}(M_Z^2)\over M^2_Z}\right)^{SUSY}
\label{eqn:dwmy}
\end{eqnarray}
In the above formulae, ~$(...)^{SUSY}$ ~stands for superparticle contributions
to the vector boson self energies and includes also the difference between
the Higgs sector contributions to these self energies in MSSM and SM.
The formulae for ~$\hat\Pi_{WW}(0)^{SUSY}$ ~and ~$\hat\Pi_{ZZ}(M_Z^2)^{SUSY}$ ~
are collected for the reader's convenience in the Appendix.

Using the interpolating formulae (\ref{eqn:fits},\ref{eqn:fitw})  for ~
$\hat s^2_{SM}$ ~and ~$(M_W)_{SM}$ ~we can check that equations
(\ref{eqn:damy},\ref{eqn:dsmy}) give an excellent approximation
to our complete results for the couplings. This is illustrated in Figs. 2-4.
The approximation for  ~$(M_W)_{MSSM}$ ~given by eq.(\ref{eqn:dwmy})
is typically within 20 MeV as compared with the full calculation
in the ~$\overline{MS}$ ~scheme.

One should stress that the FG approximation is consistent only if the
SM couplings are determined from the same set of
observables ~${\cal O}_i$ ~as used to calculate the couplings in the MSSM: ~
$g^{MSSM}_k = g^{SM}_k + \delta g_k$.  ~This is because the corrections ~
$\delta g_k$ ~are process dependent and correlated with the values of ~
$g_k^{SM}$ ~extracted from the same set of observables: since we are
concerned with the situation when the SM (without higher dimension
non--renormalizable operators) is not a sufficiently accurate effective
theory at ~$M_Z$ ~scale, the values of the couplings ~$g_k^{SM}$ ~
extracted from different experiments (a different set of observables ~
${\cal O}_i$) ~are also process dependent and ~
$g^{SM}_k({\cal O}_i) - g^{SM}_k({\cal O}_j)$ ~can be of the order of ~
$\delta g_k$ ~itself. It is therefore inconsistent to apply the FG
approach to ~$\hat s^2_{SM}$ ~obtained from an overall fit to the data
in the SM; such a fit gives meaningless numbers if
non-logarithmic SUSY corrections
are important. It is also inconsistent to apply FG method to correct
the SM couplings obtained from a set ~${\cal O}_i$ ~by using the
equations for ~$\delta g_k$ ~derived by referring to a set ~${\cal O}_j$. ~
This is shown in Figs.2-4 where the results obtained from the
two variants of the FG formulae \cite{FAGRI}
derived by reffering to the scattering processes and
applied to the SM couplings obtained from our set of observables ~
$M_Z$, ~$G_F$ and ~$\alpha_{EM}^{OS}$ ~
are marked by dotted lines.

It is easy to check, using formulae (21) that all the approximations
give the same asymptotic behaviuor of
the corrections to ~$\hat\alpha$ ~and ~$\hat s^2$ ~when the SUSY
particles are heavy. Nevertheless, as is clear from Figs. 2-4,
for light SUSY particles they give very different results from the complete
calculation.

\vskip 0.5cm
{\bf 4. UNIFICATION OF COUPLINGS.}
\vskip 0.3cm

Our calculation of the running Weinberg angle in the MSSM can be used
to investigate unification of the gauge couplings, with the supersymmetric
threshold effects taken exactly into account at the 1--loop level. The
main uncertainty in our value of ~$\sin^2\theta_W(M_Z)$ ~(for fixed~
$m_t$, ~$M_{h^0}$ ~and superpartner masses) follows from the uncertainty
in ~$\hat\alpha$ ~which gives ~$\delta\hat s^2 = \pm3\times10^{-4}$~
(theoretical uncertainity of the calculation like higher order corrections
etc. is smaller,
 ~${\cal O}(1\times10^{-4})$). ~This can be compared
with the precision ~$\pm6\times10^{-4}$ ~of ~
$\sin^2\theta_W(M_Z)$ ~in the SM obtained from the global fits to the LEP
and the low energy data \cite{LL,LP}.
Note, however, that as explained earlier, from the point of view of the MSSM
the value of the  ~$\sin^2\theta_W(M_Z)$ ~in the SM obtained from global
fits is meaningless as soon as we want to go beyond the LLT approximation.
To increase the statistical significance of the
value of ~$\hat s^2$ ~in the MSSM,
i.e. to obtain it from a global fit to more observables than just  ~
$G_F$, ~$M_Z$ and ~$\alpha_{EM}^{OS}$ ~one has to
perform similar calculation to
ours for  other observables and to have an independent fit for each
chosen set of values for ~$m_t$ ~and ~$M_{h^0}$ ~and for superpartner masses.
In particular it is important to remember about the correlation between the
value of ~$\sin^2\theta_W(M_Z)$ ~and the values of the top quark and Higgs
boson masses which is already present in the Standard Model results (see
e.g. ref. \cite{HOLL} and eq. (\ref{eqn:fits})).

Given the high accuracy of our approximate formulae
(\ref{eqn:damy},\ref{eqn:dsmy},\ref{eqn:dwmy}) which
translate ~$\hat s^2_{SM}$~ into ~$\hat s^2$~ in the
MSSM (for any chosen superpartner spectrum) and hoping
for similar accuracy of the formula connecting ~$\alpha_3^{SM}$ ~
with ~$\hat\alpha_3$ ~in the MSSM \cite{FAGRI}:
\begin{eqnarray}
{\delta\hat\alpha_3\over\hat\alpha_3^{SM}} =
-{\cal R}e \left({\hat\Pi_{gg}(M_Z)^{SUSY}\over M_Z^2}\right)
\label{eqn:dal3}
\end{eqnarray}
where ~$\hat\Pi_{gg}^{SUSY}$ ~is the SUSY contribution to the gluon self
energy (provided we apply this formula to  ~$\hat\alpha_3^{SM}$ ~
extracted from processes where the gluon four-momentum $q^2$~is of order of
$M_Z^2$), ~it is convenient to use the values of ~$\hat s^2_{SM}$ ~
and ~$\hat\alpha_3^{SM}$ ~ as a label for the set of values of ~$\hat s^2$ ~
and ~$\hat\alpha_3$ ~in the MSSM which correspond to the same values of ~
$\hat s^2_{SM}$ ~and ~$\hat\alpha_3^{SM}$ ~in the approximation
(\ref{eqn:damy},\ref{eqn:dsmy},\ref{eqn:dwmy},\ref{eqn:dal3}).
The values used in the following are as in Table 1:
\begin{eqnarray}
m_t = 160 ~{\rm GeV} ~~~ \hat s^2_{SM} = 0.23155 - 0.23195 \nonumber \\
m_t = 180 ~{\rm GeV} ~~~ \hat s^2_{SM} = 0.23093 - 0.23132 \nonumber
\end{eqnarray}
For the strong
coupling constant ~$\hat\alpha_3^{SM}$ ~ we shall be  using values ~0.115, ~
0.125  ~and ~0.13, ~for the purpose of illustration of various effects.
Thus, in this paper, by ``experimental
data'' we mean the above values of  ~$\hat s^2_{SM}$, ~$\hat\alpha_3^{SM}$ ~
and the corresponding sets of values of ~$\hat s^2$, ~$\hat\alpha_3$ ~ in
the MSSM (now dependent on SUSY parameters). Exactly analogous remarks
apply to ~$\hat\alpha^{SM}$ ~and ~$\hat\alpha$. ~
The ``experimental data'' for the MSSM will be compared with
the relation between ~$\hat s^2$, ~$\hat\alpha$ ~and ~
$\hat\alpha_3$ ~predicted after imposing the unification
conditions in the MSSM. Note that since we work directly in the MSSM,
the predictions which follow from unification can be obtained,
as a relation  between ~$\hat s^2$, ~$\hat\alpha_3$ ~and ~
$\hat\alpha$ ~without any reference to the ``experimental data''.

The final point we have to specify before the actual discussion of unification
is the choice of the supersymmetric spectrum. The very general scan over  the
sparticle masses performed in Section 3 is not very
illuminating from the point
of view of unification: by stretching the parameters and correlating them in
some specific way one has very little constraint on unification scenarios. It
is much more interesting to study unification of the gauge couplings under the
additional assumption that the supersymmetric spectrum is obtained from the
requirement of radiative electroweak symmetry breaking. In this paper we use
the spectra of ref. \cite{OLPOK} obtained with universal boundary conditions
at the GUT scale and with one -- loop corrections to the effective
potential included.  The spectra
of ref. \cite{OLPOK} correspond to a complete
scanning over the available parameter space with cut--off
on the physical squark masses ~$M_{\tilde q} < 2$ TeV. ~The
requirement of radiative breaking results in very strong
correlation between masses of different sparticles.

{}From the point  of view of the gauge couplings unification a useful
parameter which characterizes the spectrum is ~$T_{SUSY}$ ~defined so that
in the LLT approximation (neglecting GUT threshold corrections)
the SUSY threshold correction to the value of ~$\hat\alpha_3^{SM}$~
predicted from the gauge coupling unification reads \cite{LP,CARENA}:
\begin{eqnarray}
\Delta (1/\hat\alpha_3) = {1\over2\pi} {19 \over 14} \log { T_{SUSY}
\over M_Z}
\end{eqnarray}
Assuming that soft SUSY breaking mass terms are the same for all generations
we  get (see the Appendix for notation):
\begin{eqnarray}
T_{SUSY} = |\mu |
 \left(M_{g2}^2\over M^2_{\tilde g}\right)^{14\over 19}
 \left(m^2_L\over m_Q^2\right)^{3\over38}
 \left(M_{A^0}^2\over\mu^2\right)^{3\over38}
\left(M^2_{g2}\over\mu^2\right)^{2\over 19}
\end{eqnarray}

One of the consequences of the correlations
in the sparticle spectra obtained in the model with radiative
electroweak breaking
is that, generally, the effective parameter
$T_{SUSY} < 300$ GeV  ~for ~$M_{\tilde q} < 2$ TeV, ~and it is
strongly correlated with the supersymmetric parameter ~$\mu$ ~
(only in the fixed point regime, ~$T_{SUSY}$ ~extends to 600 GeV for ~
$M_{\tilde q} < 2$ TeV ~and this is due to large values of ~$\mu$, ~
which are in this case necessary for proper radiative electroweak symmetry
breaking) \cite{CARENA,COPW}.
This is shown in Fig.5  for several values of ~$m_t$
\footnote{The mass ~$m_t$~ is the pole mass which is related
to the running mass as follows: ~
\[
m_t=m_t(m_t) \left( 1+ {4 \over 3} { \alpha_3 \over \pi}
  + 11.4 \left({ \alpha_3 \over \pi}\right)^2 \right)
\] ~
In this paper, as in ref. \cite{OLPOK}, we neglect other
contributions to the pole mass.}
and ~$\tan\beta$. ~
It is important to realize that, in the context of radiative
electroweak symmetry
breaking with universal boundary conditions at the GUT scale, the often
referred to value of ~$T_{SUSY}=1$ TeV ~would correspond to  very
heavy squark spectra.

We are now equipped to study unification of the gauge couplings. First, we
address the question of unification with no GUT threshold and higher
dimension operator GUT corrections.

This requirement, as discussed before, determines a relation between ~
$\hat s^2$, ~$\hat\alpha_3$ ~and ~$\hat\alpha$ ~(all three are MSSM couplings).
Fortunately the ``experimental value'' of ~$\hat\alpha$ ~changes only weakly
for the spectra of ref. \cite{OLPOK}: ~
$(\hat\alpha_{min (max)})^{-1}$ ~= 131.8 (129.0) for heavy (light) charged
sparticles. Therefore we present the unification prediction (obtained
{}from 2-loop running with the MSSM ~RGEs from ~$M_Z$ ~to the GUT scale) as
a narrow band in the ~$\hat s^2 - \hat\alpha_3$ ~plane obtained
for ~$\hat\alpha$ ~in its ``experimental range''.
This band can be compared with ``experimental data''
for  ~$\hat s^2$ ~and ~$\hat\alpha_3$ ~ obtained with the use of the chosen
SUSY spectra, bearing in mind that  ~$(\hat\alpha_{min (max)})^{-1}$~ is more
appropriate for heavy (light) spectra.
Labelling the ``experimental data'' by the
corresponding Standard Model values of ~$\hat s^2_{SM}$ ~
and ~$\hat\alpha_3^{SM}$ ~we see in Fig. 6 a clear trend: unification with no
GUT threshold corrections occurs only for sufficiently large ~
$\hat\alpha^{SM}_3$ ~and sufficiently heavy supersymmetric spectrum
(the actual numbers depend on ~$m_t$ ~and ~$\tan\beta$ ~values).
The dependence on ~$m_t$~ has two sources: firstly
it enters into
the 1-loop correction to ~$\hat s^2$ ~and also to ~$\hat\alpha_3$
\footnote{The coupling constant usually quoted in the literature which
          we call ~$\hat\alpha_3^{SM}$ ~throughout the paper is actually ~
          $\hat\alpha_3^{QCD+QED}$~in the theory with the
          top quark decoupled.} ~
and ~$\hat\alpha$~ as
already in the SM (see the values (\ref{eqn:fits})).
Secondly the features of the SUSY spectrum depend
on ~$m_t$ ~and ~$\tan\beta$ ~because of the tight constraints imposed
by universality of the soft terms and by radiative symmetry breaking.
In particular, as discussed above and seen in Fig. 5 for
the same cut -- off for ~$M_{\tilde q}$ ~the maximal value of ~
$T_{SUSY}$ ~is ~$m_t$ ~and ~$\tan\beta$ ~dependent.
In Fig. 7 we plot the predicted ~$\hat\alpha_3^{SM}$ ~
in the approximation (\ref{eqn:damy},\ref{eqn:dsmy},\ref{eqn:dal3})
(from unification with no GUT thresholds)
as a function of ~$T_{SUSY}$ ~for several values of ~$m_t$ ~and ~$\tan\beta$ ~
and with ~$\hat s^2_{SM}$ ~and ~
$\hat\alpha_{SM}$ ~in their experimental ranges specified earlier.
The uncertainty ~$\delta\hat s^2 = \pm3\times10^{-4}$~
(correlated with uncertainty in $\hat\alpha$,
eq. (\ref{eqn:dalfa})) translates itself into ~
$\delta\hat \alpha_3^{SM} = \mp1.5\times10^{-3}$~
in this plot. The dependence of ~$\hat s^2_{SM}$ ~on ~$m_t$ ~
(eq. (\ref{eqn:fits})) is of
course reflected in the predicted value of ~$\hat\alpha_3^{SM}$. ~
Changing ~$m_t$ ~from 160 GeV to $160 \pm 20$~ GeV
(and leaving the sparticle spectrum unaffected)
results in a shift
$\delta_t\hat\alpha_3^{SM} = \pm(2 - 2.5)\times10^{-3}$~.

In the LLT approximation and neglecting the dependence of ~$\hat s^2_{SM}$ ~
on the Higgs boson mass, the predicted ~$\hat\alpha_3^{SM}$ ~is a
function of ~$T_{SUSY}$ ~only. The scatter plot obtained in the LLT
approximation reflects the dependence of the input ~$\hat s^2_{SM}$ ~
on the Higgs boson mass which is varied from 60 to 130 GeV. ~In the
full calculation there is stronger departure from a universal
dependence of ~$\hat\alpha_3^{SM}$ ~on ~$T_{SUSY}$ ~which is
due to the non-logarithmic threshold corrections.
Going beyond the LLT approximation results in
larger values of ~$\hat\alpha_3^{SM}$ ~for small ~$T_{SUSY}$. ~
Moreover, in the whole range of ~$T_{SUSY}$ ~reachable with our spectra,
the corrections to the LLT approximation are non--negligible.
This happens because the spectra always contain some
light sparticle(s) which is responsible for the difference.
In the absence of the GUT threshold corrections the unification  scale
can be unambigously defined as the point of crossover of the couplings.
It depends only  weakly on the SUSY spectra and varies
between ~$1.3\times 10^{16}$ ~and ~$3\times 10^{16}$.

Next we study unification with GUT thresholds included. We restrict
ourselves to the minimal supersymmetric ~$SU(5)$ ~with three different mass
scales: ~$M_{H_3}$, ~$M_{\Sigma}$ ~and ~$M_V$ ~corresponding to the Higgs
triplet, Higgs bosons originating from the ~$\underline {24}$ ~representation
and heavy  vector bosons (leptoquarks) respectively.
We have then \cite{JAP,BH}:
\begin{eqnarray}
{1\over\alpha_3(Q)} = {1\over\alpha_5(Q)}
+ {1\over2\pi}\left(4\log{M_V\over Q} - 3\log{M_{\Sigma}\over Q}
- \log{M_{H_3}\over Q}\right)\nonumber
\end{eqnarray}
\begin{eqnarray}
{1\over\alpha_2(Q)} = {1\over\alpha_5(Q)}
+ {1\over2\pi}\left(6\log{M_V\over Q} - 2\log{M_{\Sigma}\over Q}\right)
\label{eqn:thrgut}
\end{eqnarray}
\begin{eqnarray}
{1\over\alpha_1(Q)} = {1\over\alpha_5(Q)}
+ {1\over2\pi}\left(10\log{M_V\over Q} - {2\over5}\log{M_{H_3}\over Q}
\right)\nonumber
\end{eqnarray}
for any scale ~$Q$ ~close to ~$M_{V,\Sigma,H_3}$ ~i.e. a scale
at which the higher order (logarithmic) threshold corrections to the RHS
can be neglected.
It follows from eqs. (\ref{eqn:thrgut}) that ~$M_{H_3}$, ~and
a combination of $M_{\Sigma}$ ~and ~$M_V$ ~are uniquely related to the
values of the coupling constants:
\begin{eqnarray}
\left({3\over\alpha_2}-{2\over\alpha_3}-{1\over\alpha_1}\right)(Q) &=&
{6\over5\pi}\log{M_{H_3}\over Q}\nonumber\\
\left({5\over\alpha_1}-{3\over\alpha_2}-{2\over\alpha_3}\right)(Q) &=&
{6\over\pi}\log{M_V^2M_{\Sigma}\over Q^3}
\label{eqn:mgut}
\end{eqnarray}
and ~$\alpha_i(Q)$ ~are related to ~$\hat\alpha_i\equiv\alpha_i(M_Z)$ ~
by the RGE in the MSSM.

The curves of fixed ~$M_{H_3}$ ~in the ~$(\hat s^2, \hat\alpha_3)$ ~
plane are given by eq. (\ref{eqn:mgut}) (for  fixed ~$\hat\alpha_{EM}$). ~
They are shown in Fig.6 for ~$M_{H_3}= 5\times10^{15}, ~2\times10^{16}, ~
1\times10^{17}$ ~(for ~$\hat\alpha_{EM}=1/131.8$). ~
Thus, each point in the ~$(\hat s^2, \hat\alpha_3)$ ~
plane  corresponds to fixed values of ~$M_{H_3}$ ~and of the product ~
$M_V^2M_{\Sigma}$ ~(modulo the uncertainity in ~$\hat\alpha_{EM}$).
Running with 2-loop equations from ~$M_Z$ ~to Q such that
LHS of the second of eqs. (\ref{eqn:mgut}) vanishes we obtain
$(M_V^2 M_{\Sigma})^{1/3} ~ (=Q)$ ~and $M_{H_3}$.
With GUT threshold corrections, unification of couplings can
be always achieved at the expense of large enough
splitting between ~
$M_{H_3}$
and the combination
$(M_V^2 M_{\Sigma})^{1/3}$.
There are, however,
certain natural requirements which constrain
these three masses and their splitting. First of all,
$M_{H_3}$, ~$M_{\Sigma}$ ~and ~$M_V$ ~should be smaller than the Planck mass.
Other requirements follow from the superpotential which breaks ~$SU(5)$~
and the condition that its couplings remain perturbative up to the Planck
scale. This gives \cite{JAP}:
\begin{eqnarray}
M_{H_3} < 2  ~M_V,  ~~~~~~~~~M_{\Sigma} < 1.8 ~M_V
\label{eqn:gutbounds1}
\end{eqnarray}
Another requirement is more technical: the discussion of unification based
on the 1--loop threshold corrections eqs. (\ref{eqn:thrgut}) is
meaningful only if there exist a common scale ~$Q$ ~such that all logarithms~
$\log(M_{H_3}/Q)$, ~$\log(M_{\Sigma}/Q)$ ~and ~$\log(M_V/Q)$, ~are small
enough to make higher order corrections (next to leading logarithms)
negligible. Those arguments still leave a lot of freedom for GUT
corrections.
It is interesting to know how much, say,
the predicted ~$\hat\alpha_3^{SM}$~ is modified by the splitting ~
$(M_V^2 M_\Sigma)^{1/3}/M_{H_3}>>1$. ~
Using the formulae (\ref{eqn:mgut}) with 1-loop
running to ~$Q=M_Z$ ~we obtain
\begin{eqnarray}
\delta_{GUT} \left({1\over\hat\alpha_3}\right) =
-{3\over{14\pi}}\log{M_{H_3}^3 \over M_V^2M_{\Sigma}}
\label{eqn:delgut}
\end{eqnarray}
which yields ~$\delta_{GUT}\hat\alpha_3^{SM}= -0.01~( -0.02)$ ~for ~
$(M_V^2 M_\Sigma)^{1/3}/M_{H_3}=30~(1000)$ ~
(which is not strongly modified by 2--loop effects)
respectively.
We demonstrate the effect of GUT scale threshold correction
in Fig. 6 where we show the bounds in
the ~$(\hat s^2, \hat\alpha_3)$ ~plane in the MSSM which are consistent with
unification with
\begin{eqnarray}
{max(M_{H_3}, ~M_{\Sigma}, ~M_V)\over min(M_{H_3}, ~M_{\Sigma}, ~M_V)} <
30 ~{\rm or} ~1000
\label{eqn:gutbounds2}
\end{eqnarray}
respectively. The only (weak) dependence of those bounds on the supersymmetric
spectrum is  through the value of ~$\hat\alpha_{EM}$. ~
The asymmetry of the bounds with respects to the unification curve with
no GUT threshold corrections, seen in Fig.6, is due to condition
(\ref{eqn:gutbounds1}). Below this curve  ~$M_V=M_{\Sigma}=30 ~(1000)~
M_{H_3}$~ and above ~$M_{H_3}=2 M_V=30 ~(1000)~M_{\Sigma}$ which
results in a much smaller splitting between ~$M_{H_3}$ ~and ~
$(M_V^2 M_\Sigma)^{1/3}$. ~

As stressed in ref. \cite{BH}, the uncertainty in the
unification prediction for ~$\hat\alpha_3$ ~due to the unknown
GUT thresholds makes direct tests of the unification idea
impossible, even if we knew the low energy superparticle
spectrum and no matter how precise the experimental data for the
couplings at ~$M_Z$ ~are. Only the reversed programme is
realizable: the more precise the experimental data for the couplings
are the more accurate predictions we can obtain for ~$M_{H_3}$ ~and
for the splitting between ~$M_{H_3}$~ and ~$(M_V^2 M_\Sigma)^{1/3}$. ~
Comparing the bounds with our ``experimental data'' it is clear
that for light SUSY spectrum GUT threshold corrections, corresponding
to large splitting ~$M_{V(\Sigma)}/M_{H_3}\sim30$ ~are needed for unification
with ~$\hat\alpha_3^{SM}=0.125$. ~If one takes ~$\hat\alpha_3^{SM}=0.115$, ~
close to its experimental lower bound, the same GUT splitting is needed
even for ~$T_{SUSY}=300$ GeV ~and it has to reach  ~
$M_{V(\Sigma)}/M_{H_3}\sim1000$ ~for light spectra.
For the same value of $\hat\alpha_3^{SM}$, the non-logarithmic corrections
calculated in this paper tend to increase GUT corrections needed to achieve
unification.

It is also very interesting to show the predictions for ~$M_{H_3}$ ~which
follow from our ``experimental data'' points. The Higgs triplet mass is
directly related to the proton lifetime \cite{PROTON}. There have been
claims in the literature
that even ~$M_{H_3}\sim 10^{17}$ GeV  can be reconciled
with the gauge coupling unification and at the same time it is large enough to
make the proton life time consistent with experimental limits \cite{PROTONEXP}
also for large values of ~$\tan\beta$ ~\cite{JAP}. Our predictions for ~
$M_{H_3}$ ~ are plotted in Fig. 8 as a function of ~$T_{SUSY}$ ~
for ~$\hat\alpha_3^{SM}=0.125$. ~The value of ~$M_{H_3}$~ is rather sensitive
to changes of the MSSM couplings at ~$M_Z$. ~For example, a change ~
$\delta\hat s^2 = 3\times10^{-4}$ ~increases ~$M_{H_3}$~ by a
factor of ~$\sim$1.5 ~
whereas a change ~$\delta\hat\alpha_3 = -2\times10^{-3}$ ~decreases it
by a factor ~$\sim$2.2.

We stress
again that those results depend on the SUSY spectrum which is taken from
the minimal supergravity model  (radiative electroweak symmetry
breaking with universal boundary conditions at the GUT scale) and with
cut--off on the physical squark masses, ~$M_{\tilde q} <2$ TeV. ~
We obtain an upper bound on ~$M_{H_3}$~ which is lower
than the one in ref. \cite{JAP}. For the same value
of ~$\mu=1$ TeV, ~$\hat\alpha_3^{SM}=0.125, ~m_t=160$ GeV ~and ~
including the uncertainty
$\delta\hat s^2 =\pm3\times10^{-4}$ ~we
obtain ~$M_{H_3}^{max}=3\times 10^{16}$~.
The discrepancy with the upper bound given in ref. \cite{JAP} is
due to a different $s^2_{SM}$~ value used in that paper.
Our bounds on ~$M_{H_3}$~are in agreement with those given in ref.
\cite{NATH}. We also observe that our non--logarithmic SUSY threshold
corrections decrease the predicted value of ~$M_{H_3}$. ~

Finally we would like to discuss the impact on the unification of gauge
couplings of the condition that quark Yukawa couplings remain in the
perturbative regime. In particular we impose
\begin{eqnarray}
Y^2_t < 4\pi
\label{eqn:yukbound}
\end{eqnarray}
This condition is relevant only for a heavy top quark and small ~$\tan\beta$~
(i.e. close to the quasi IR--fixed point \cite{FIXED,CARENA}).
As it is clear from the
structure of the RGE for the ~$Y_t$, ~ for given values of ~$m_t$ ~
and ~$\tan\beta$, ~the condition (\ref{eqn:yukbound}) gives a lower
bound on ~$\hat\alpha_3^{MSSM}$
\footnote{The bound  depends (weakly) on ~$\sin^2\theta_W$ ~because
          a change in ~$\sin^2\theta_W$ ~ induces a change in the final
          point of the renormalization evolution where ~$Y_{top}$ ~grows
          very quickly. In this discussion  we neglect the contribution of SUSY
          and GUT threshold corrections to the Yukawa couplings and in this
          sense our conclusions here should be taken as only qualitative.}. ~
Superimposed on the unification plots in Fig. 6, it favours light superpartner
spectrum: for heavy spectrum the values of ~$\hat\alpha_3^{MSSM}$ ~
corresponding to resonable values of ~$\hat\alpha_3^{SM}$ ~(say,
below 0.13) are too small to be consistent with perturbative top quark
Yukawa coupling. It is interesting to observe that, for small ~$\tan\beta$, ~
large ~$m_t$, ~and ~$\hat\alpha_3^{SM}<
0.125(0.115)$, ~perturbativity
of ~$Y_t$ ~puts strong upper
bounds on ~$T_{SUSY}$: ~$T_{SUSY} < 200 (~600)$ GeV  ~for ~
$m_t = 160$ (180) GeV ~and  ~$\tan\beta= 1.25~(2.0)$ ~respectively.

\vskip 0.5cm

{\bf 5. CONCLUSIONS.}
\vskip 0.3cm

In this paper we have calculated the running Weinberg angle at ~$M_Z$ ~
directly in the MSSM in terms of ~$G_F$, ~$M_Z$, ~$\alpha_{EM}^{OS}$, ~
$m_t$ ~and supersymmetric parameters. The accuracy of this calculation
is similar to the analogous calculation in the Standard Model. Using the
language of supersymmetric threshold corrections to the Standard Model
our calculation is equivalent to a complete calculation of those corrections
at the one -- loop level. A detailed comparison with LLT approximation is
presented. Also highly accurate approximate formulae are given which
translate the SM couplings into the MSSM couplings.

Those results are subsequently used to study the gauge coupling unification
in the minimal supersymmetric ~$SU(5)$ ~model, with the superpartner
spectra taken from the minimal supergravity model (i.e. with radiative
electroweak symmetry breaking and universal boundary conditions for soft
supersymmetry breaking parameters at the unification scale).

For unification with no GUT thresholds, the value of the strong coupling
constant is calculated as a function of supersymmetric spectra.
The non-logarithmic threshold corrections increase the predicted
value of ~$\hat\alpha_3^{SM}$ ~for light spectra. For the spectra obtained
in the minimal supergravity model with radiative electroweak breaking,
with cut--off ~$M_{\tilde q} <2$ TeV, ~one generically obtains ~
$\hat\alpha_3^{SM}$~ above the experimental range  ~
$\hat\alpha_3^{SM}= 0.118 \pm 0.007$ ~(with the exception
of the low ~$\tan\beta$ ~values and ~$m_t$ ~close to its quasi--IR fixed
point).
In the presence of GUT thresholds, a direct precise
test of the unification idea in the minimal SU(5) is no longer possible
\cite{BH}. Only the reversed programme is realizable: with precise
information on the
couplings at ~$M_Z$~ we can calculate the Higgs triplet mass and
the combination ~$(M_V^2 M_\Sigma)^{1/3}$ ~as a function of the
supersymmetric spectrum. For the same values of the couplings at ~$M_Z$, ~
the non-logarithmic corrections calculated in this paper decrease
the value of ~$M_{H_3}$ ~and increase the splitting between  ~
$M_{H_3}$~ and ~$(M_V^2 M_\Sigma)^{1/3}$~ needed to achieve
unification. Information from the proton
life time (sensitive to ~$M_{H_3}$) ~and from ~$b-\tau$ ~Yukawa coupling
unification (sensitive to the splitting) gives additional constraints on
the model. It will be interesting to use the results of the present paper for
improving the precision of those constraints.

\vskip 0.5cm
\noindent {\bf Acknowledgments.}
P.H. Ch. would like to thank Professor W.F.L. Hollik for his kind
hospitality extended to him during his stay at the University of Karlsruhe
and many discussions. His work was partly supported
by European Union under contract CHRX-CT92-0004.
Z.P. Is grateful to M. Olechowski for providing him with programs
computing MSSM spectra and many disscussions.
S.P. is grateful to the Aspen Center for Physics for its hospitality
during the completion of this work.

\newpage
{\bf APPENDIX.}
\vskip 0.3cm

In this Appendix we make explicit our convention and notation for
sparticle masses. In order to avoid confusion
with signs, often present in the literature, we will be explicit here.
We follow closely the convention (but not the notation) of \cite{ROS}.

{}From the superpotential of the model
\begin{eqnarray}
{\cal W} = \epsilon_{ij}Y_l\hat H^i_1\hat L^j\hat E^+_R
+ \epsilon_{ij}Y_d\hat H^i_1\hat Q^j\hat D^+_R
+ \epsilon_{ij}Y_u\hat H^i_2\hat Q^j\hat U^-_R
+ \epsilon_{ij}\mu\hat H^i_1\hat H^j_2
\end{eqnarray}
(where ~$Y_a$ ~are the Yukawa couplings and~
$\epsilon_{12}=-\epsilon_{21}=-1$) ~and the relevant soft SUSY breaking
part of the lagrangian
\begin{eqnarray}
{\cal L}_{soft} &=& -~m^2_{H_1} |H_1|^2 - m^2_{H_2} |H_2|^2
+ m^2_3 \epsilon_{ij} (H^i_1H^j_2 + c.c.) \nonumber\\
&-&m^2_L |L|^2 - m^2_E |E^+_R|^2
-m^2_Q |Q|^2 - m^2_U |U^-_R|^2 - m^2_D  |D^+_R|^2\\
&+& \epsilon_{ij}Y_lA_lH^i_1L^jE^+_R
+ \epsilon_{ij}Y_dA_dH^i_1Q^jD^+_R + \epsilon_{ij}Y_lA_lH^i_2Q^jU^-_R
\nonumber
\end{eqnarray}
where ~$H_1$, ~$H_2$ ~and ~$L$, ~$Q$ ~are the ~$SU(2)$ ~doublets and
adding contributions of the D--terms one gets the mass matrices  for
up and down squarks as well as charged sleptons
(because we neglect the intergenerational mixing
we write them for one generation only; therefore ~$U ~(u)$ ~represents
generically up--type squarks (quarks) and the same is understood
for dow--type ones and sleptons (leptons)):
\begin{eqnarray}
{\cal M}^2_U = \left(\matrix{
M^2_{U_L} & M^2_{U_{LR}}\cr
M^2_{U_{LR}} & M^2_{U_R} }
\right)
\end{eqnarray}
with
\begin{eqnarray}
M^2_{U_L} &=& m^2_Q + m^2_u + {1\over6}t(M^2_Z-4M^2_W)\nonumber \\
M^2_{U_R} &=& m^2_U + m^2_u - {2\over3}t (M^2_Z-M^2_W)\\
M^2_{U_LR} &=& -m_u(A_u + \mu\cot\beta)\nonumber
\end{eqnarray}
\begin{eqnarray}
M^2_{D_L} &=& m^2_Q + m^2_u - {1\over6}t(M^2_Z+2M^2_W)\nonumber \\
M^2_{D_R} &=& m^2_D + m^2_u + {1\over3}t (M^2_Z-M^2_W)\\
M^2_{D_LR} &=& -m_d(A_d + \mu\tan\beta)\nonumber
\end{eqnarray}
\begin{eqnarray}
M^2_{E_L} &=& m^2_L + m^2_e - {1\over2}t(M^2_Z-2M^2_W)\nonumber \\
M^2_{E_R} &=& m^2_E + m^2_e + t (M^2_Z-M^2_W)\\
M^2_{E_LR} &=& -m_e(A_e + \mu\tan\beta)\nonumber
\end{eqnarray}
Here ~$t\equiv (\tan^2\beta-1)/(\tan^2\beta+1)$ ~and ~
$m_{u,d,e}$ ~stand for ordinary fermion masses.
These matrices are diagonalized by the appropriate rotations.
The sneutrinos have masses given by:
\begin{eqnarray}
M^2_{\tilde\nu} = m^2_L + m^2_e + {t\over2}  M^2_Z
\end{eqnarray}

Part of the lagrangian relevant for the chargino/neutralino sector reads:
\begin{eqnarray}
{\cal L}&=&{1\over2}M_{g2}\psi^a\psi^a + {1\over2}M_{g1}\chi\chi
-\mu\epsilon_{ij}h^i_1h^j_2+i\sqrt2 g_2 H^*_1 T^a h_1 \psi^a \\
&+&i\sqrt2 g_2 H^*_2 T^a h_2 \psi^a
-{i\over\sqrt2} g_y H^*_1 h_1 \chi +{i\over\sqrt2} g_y H^*_2 h_2 \chi
+ h.c.\nonumber
\end{eqnarray}
where ~$\psi^a$ ~(a=1,2,3) and ~$\chi$ ~are the ~$SU(2)$ ~and ~$U(1)$ ~
gauginos respectively, ~$h_1$ ~and ~$h_2$ ~are the two higgsino doublets
and ~$T^a$ ~are the ~$SU(2)$ generators.
Definig ~$\sqrt2\psi^{\pm}=\psi^1\mp i \psi^2$ ~we get in the basis~
$(-i\psi^+, h^1_2,-i\psi^-,h^2_1)\equiv (\chi^+,\chi^-)$ ~the chargino mass
matrix:
\begin{eqnarray}
{\cal L}_{mass} = -{1\over2} (\chi^+,\chi^-)
\left(\matrix{0 & X^T\cr X & 0}\right) \left(\matrix{\chi^+\cr\chi^-}\right)
+ h.c.
\end{eqnarray}
with
\begin{eqnarray}
X=\left(\matrix{M_{g2}&\sqrt2M_W\sin\beta\cr\sqrt2M_W\cos\beta &\mu}\right)
\end{eqnarray}
Diagonalization requires two unitary mixing matrices ~$Z_+$ ~and ~$Z_-$
which rotate ~$\chi^{\pm}$ ~fields to mass eigenstates ~$\lambda^{\pm}$:~
$\lambda^{\pm} = Z^{\dagger}_{\pm}\chi^{\pm}$. The two physical chargino
masses are denoted as ~$m_{C_i}$, ~$i=1,2$.

The neutralino mass matrix in the basis ~
$\chi^0 = (-i\chi,-i\psi^3,h^1_1,h^2_2)$ ~has the well known form:
\begin{eqnarray}
{\cal M}_N = \left(\matrix{
M_{g1} & 0 & -M_Z \hat s\cos\beta & M_Z \hat s\sin\beta \cr
0 & M_{g2} &  M_Z \hat c\cos\beta & -M_Z \hat c\cos\beta \cr
-M_Z \hat s\cos\beta & M_Z \hat c\cos\beta & 0 & -\mu \cr
M_Z \hat s\sin\beta & -M_Z \hat c\cos\beta & -\mu & 0}\right)
\end{eqnarray}
and is diagonalized by rotation ~$\lambda^0 = Z_N^{-1} \chi^0$. ~
Physical neutralino masses are denoted as ~$m_{N_i}$, ~$i=1,...4$. ~Four
component chargino (Dirac) and neutralino (Majorana) fields are then built as
\begin{eqnarray}
\psi_{C^+_i}=\left(\matrix{\lambda^+_i\cr\overline\lambda^-_i}\right)~~~
\psi_{N_i}=\left(\matrix{\lambda^0_i\cr\overline\lambda^0_i}\right)
\end{eqnarray}

Finally, gluino, which does not mix with anything, has a mass denoted
as ~$M_{\tilde g}$. ~

We do not describe here the Higgs sector referring
the reader to refs. \cite{EZ,MY_H}. We recall only that at the tree level,
the Higgs boson masses and couplings can be conveniently parametrized by
the ratio of the vacuum expectation values of the two Higgs doublets ~
$\tan\beta\equiv v_2/v_1$ ~and the ~$CP-$odd Higgs boson mass ~$M_{A^0}$.

In terms of the parameters defined above, the general scan used in
Sec. 3 can be described as follows. ~
$50 < M_{A^0} < 500$ GeV, ~$50 < m_{C_j}, ~m_{N_i} < 1500$ GeV,~
$120 < M_{B_L} < 1000$ GeV, ~
$0.4 M_{B_L} < M_{T_R} < 2.5  M_{T_L}$,~
$0 < A_t < M_{B_L}$, ~
$0.4M_{B_L} < M_{\tilde\nu} < 0.8 M_{B_L}$~
keeping ~$M_{Q_L} = M_{Q_R} =M_{B_L}$  ~and~
$A_q=0$ ~for the first two generations  as well as~
$M_{E_R} = M_{\tilde\nu}$ ~with ~$A_l = 0$~ (for all the
three generations).

Next, we display the formulae for ~$(\hat\Pi_{WW}(q^2))^{SUSY}$~
and ~$(\hat\Pi_{ZZ}(q^2))^{SUSY}$ ~which allow easy use of our
approximation given by eqs.
(\ref{eqn:fits},\ref{eqn:fitw},\ref{eqn:damy},\ref{eqn:dsmy},\ref{eqn:dwmy}).
As explained in Sec. 3, besides the genuine sparticle contributions, ~
$(\hat\Pi_{WW}(q^2))^{SUSY}$ ~and ~$(\hat\Pi_{ZZ}(q^2))^{SUSY}$ ~
include also the difference of contributions coming from the Higgs
sectors of MSSM and SM and therefore depend also on the SM Higgs boson mass ~
$M_{\phi^0}$. ~Obviously, calculating the corrections to ~$\hat s^2$ ~one
has to take the same ~$M_{\phi^0}$ ~in formulae below and in the fit
(\ref{eqn:fits}). It is convenient to choose ~$M_{\phi^0} = M_{h^0}$.

Let us first consider ~$W^{\pm}$ ~self energy. The difference of the MSSM and
SM  Higgs boson contributions gives:
\begin{eqnarray}
4\pi(\hat\Pi_{WW}(q^2))^{h}&=&
{\hat\alpha\over\hat s^2} M^2_W \left[b_0(q^2,M_W,M_{\phi^0})
\right.\nonumber\\
&-&\left.c^2_{\beta\alpha}b_0(q^2,M_W,M_{H^0})
-s^2_{\beta\alpha}b_0(q^2,M_W,M_{h^0})\right]\nonumber\\
&+&{\hat\alpha\over\hat s^2}\left[-\tilde b_{22}(q^2,M_W,M_{\phi^0})
+\tilde b_{22}(q^2,M_{H^+},M_{A^0})\right.\\
&+&s^2_{\beta\alpha}\tilde b_{22}(q^2,M_{H^+},M_{H^0})
+c^2_{\beta\alpha}\tilde b_{22}(q^2,M_{H^+},M_{h^0})\nonumber\\
&+&\left.c^2_{\beta\alpha}\tilde b_{22}(q^2,M_W,M_{H^0})
+s^2_{\beta\alpha}\tilde b_{22}(q^2,M_W,M_{h^0})\right]\nonumber
\end{eqnarray}
where the functions ~$b_0$ ~and ~$\tilde b_{22}$ ~are defined in
eqs.(\ref{eqn:defb22},\ref{eqn:defb0}),~
$s^2_{\beta\alpha}\equiv\sin^2(\beta-\alpha)$ ~and ~$\alpha$ ~is
the mixing angle between the two MSSM scalar Higgs bosons defined e.g. as
in \cite{HHG,ROS}. We recall that we use ~$M_{h^0}$, ~$M_{H^0}$ and ~
$\alpha$ ~with 1--loop corrections included via the effective potential
approach of ref. \cite{EZ}.

Each generation of sleptons contributes
\begin{eqnarray}
4\pi(\hat\Pi_{WW}(q^2))^{sl}=2{\hat\alpha\over\hat s^2}
\left[c^2_L\tilde b_{22}(q^2,M_{E_1},M_{\tilde\nu})
+s^2_L\tilde b_{22}(q^2,M_{E_2},M_{\tilde\nu})\right]
\end{eqnarray}
where ~$c_L\equiv\cos\phi_L$ ~and the mass eigenstates of charged sleptons
are defined by their relations to the left and right handed states:
\begin{eqnarray}
E^-_L = c_L E^-_1 - s_L E^-_2~~~E^-_R = s_L E^-_1 + c_L E^-_2
\label{eqn:mixl}
\end{eqnarray}
Similarily, one generation of squarks gives
\begin{eqnarray}
4\pi(\hat\Pi_{WW}(q^2))^{sq}&=&6{\hat\alpha\over\hat s^2}
\left[c^2_Uc^2_D\tilde b_{22}(q^2,M_{U_1},M_{D_1})
+s^2_Uc^2_D\tilde b_{22}(q^2,M_{U_2},M_{D_1})\right.\nonumber\\
&+&\left.c^2_Us^2_D\tilde b_{22}(q^2,M_{U_1},M_{D_2})
+s^2_Us^2_D\tilde b_{22}(q^2,M_{U_2},M_{D_2})\right]
\end{eqnarray}
with ~$c_{U,D}\equiv\cos\phi_{U,D}$ ~defined by
\begin{eqnarray}
U^+_L = c_U U^+_1 - s_U U^+_2~~~ U^+_R = s_U U^+_1 + c_U U^+_2\nonumber\\
D^-_L = c_D D^-_1 - s_D D^-_2~~~ D^-_R = s_D D^-_1 + c_D D^-_2
\label{eqn:mixq}
\end{eqnarray}
Finally contribution of two charginos and four neutralinos reads:
\begin{eqnarray}
4\pi(\hat\Pi_{WW}(q^2))^{CN}&=&{\hat\alpha\over\hat s^2}
\sum_{j=1}^2\sum_{i=1}^4(|c_{L,C/N}^{ji}|^2+|c_{R,C/N}^{ji}|^2)
\left[-4\tilde b_{22}(q^2,m_{C_j},m_{N_i})\right.\nonumber\\
&-&\left.(q^2 - m^2_{C_j} - m^2_{N_i}) b_0(q^2,m_{C_j},m_{N_i})\right]\\
&-&{\hat\alpha\over\hat s^2}\sum_{j=1}^2\sum_{i=1}^4
4{\cal R}e(c_{L,C/N}^{ji}c_{R,C/N}^{ji*})m_{C_j}m_{N_i}b_0(q^2,m_{C_j},m_{N_i})
\nonumber
\label{eqn:piwwnc}
\end{eqnarray}
The explicit formulae for the left and right couplings ~$c_{L,R, C/N}^{ji}$ ~
of charginos and neutralinos to ~$W^{\pm}$ ~will be given shortly.

For the ~$\hat\Pi_{ZZ}$ ~the difference of the MSSM and SM Higgs sectors reads:
\begin{eqnarray}
4\pi(\hat\Pi_{ZZ}(q^2))^{h}&=&
{\hat\alpha\over\hat s^2\hat c^2} M^2_Z\left[b_0(q^2,M_Z,M_{\phi^0})\right.
\nonumber\\
&-&\left.c^2_{\beta\alpha}b_0(q^2,M_Z,M_{H^0})
-s^2_{\beta\alpha}b_0(q^2,M_Z,M_{h^0})\right]\nonumber\\
&+&{\hat\alpha\over\hat s^2\hat c^2}\left[
s^2_{\beta\alpha}\tilde b_{22}(q^2,M_Z,M_{h^0})
+c^2_{\beta\alpha}\tilde b_{22}(q^2,M_Z,M_{H^0})\right.\\
&+&\left.c^2_{\beta\alpha}\tilde b_{22}(q^2,M_{A^0},M_{h^0})
+s^2_{\beta\alpha}\tilde b_{22}(q^2,M_{A^0},M_{H^0})\right]\nonumber\\
&+&{\hat\alpha\over\hat s^2\hat c^2}\left[
(\hat c^2-\hat s^2)^2 \tilde b_{22}(q^2,M_{H^+}, M_{H^+})
-\tilde b_{22}(q^2,M_Z,M_{\phi^0})\right]\nonumber
\end{eqnarray}
One generation of sleptons yields:
\begin{eqnarray}
4\pi(\hat\Pi_{ZZ}(q^2))^{sl}={\hat\alpha\over\hat s^2\hat c^2}
\left[\tilde b_{22}(q^2,M_{\tilde\nu},M_{\tilde\nu})
+(c^2_L-2\hat s^2)^2\tilde b_{22}(q^2,M_{E_1},M_{E_1})
\right.\nonumber\\
+\left.(s^2_L-2\hat s^2)^2\tilde b_{22}(q^2,M_{E_2},M_{E_2})
+2c^2_Ls^2_L\tilde b_{22}(q^2,M_{E_1},M_{E_2})\right]
\end{eqnarray}
Likewise one generation of squarks contributes:
\begin{eqnarray}
4\pi(\hat\Pi_{ZZ}(q^2))^{sq}=3{\hat\alpha\over\hat s^2\hat c^2}
\left[2c^2_Us^2_U\tilde b_{22}(q^2,M_{U_1},M_{U_2})
+2c^2_Ds^2_D\tilde b_{22}(q^2,M_{D_1},M_{D_2})\right.\nonumber\\
+(c^2_U-{4\over3}\hat s^2)^2\tilde b_{22}(q^2,M_{U_1},M_{U_1})
+(s^2_U-{4\over3}\hat s^2)^2\tilde b_{22}(q^2,M_{U_2},M_{U_2})\nonumber\\
+\left.(c^2_D-{2\over3}\hat s^2)^2\tilde b_{22}(q^2,M_{D_1},M_{D_1})
+(s^2_D-{2\over3}\hat s^2)^2\tilde b_{22}(q^2,M_{D_2},M_{D_2})\right]
\nonumber\\
\end{eqnarray}
Two charginos and four neutralinos give
\begin{eqnarray}
4\pi(\hat\Pi_{ZZ}(q^2))^{CN}&=&{1\over4}{\hat\alpha\over\hat s^2\hat c^2}
\sum_{j=1}^2\sum_{i=1}^2(|c_{L,C}^{ji}|^2+|c_{R,C}^{ji}|^2)
\left[-4\tilde b_{22}(q^2,m_{C_j},m_{C_i})\right.\nonumber\\
&-&\left.(q^2 - m^2_{C_j} - m^2_{C_i}) b_0(q^2,m_{C_j},m_{C_i})\right]
\nonumber\\
&-&{\hat\alpha\over\hat s^2}\sum_{j=1}^2\sum_{i=1}^2
{\cal R}e(c_{L,C}^{ji}c_{R,C}^{ji*})m_{C_j}m_{C_i}b_0(q^2,m_{C_j},m_{C_i})
\nonumber\\
&+&{1\over8}{\hat\alpha\over\hat s^2\hat c^2}
\sum_{j=1}^4\sum_{i=1}^4(|c_{L,N}^{ji}|^2+|c_{R,N}^{ji}|^2)
\left[-4\tilde b_{22}(q^2,m_{N_j},m_{N_i})\right.\nonumber\\
&-&\left.(q^2 - m^2_{N_j} - m^2_{N_i}) b_0(q^2,m_{N_j},m_{N_i})\right]\\
&-&{1\over2}{\hat\alpha\over\hat s^2}\sum_{j=1}^4\sum_{i=1}^4
{\cal R}e(c_{L,N}^{ji}c_{R,N}^{ji*})m_{N_j}m_{N_i}b_0(q^2,m_{N_j},m_{N_i})
\nonumber
\label{eqn:pizznc}
\end{eqnarray}

The function ~$\tilde b_{22}$ ~can be easily expressed in terms of standard ~
$a$ ~and ~$b_0$ ~functions \cite{PV,MY_H} as:
\begin{eqnarray}
\tilde b_{22}(q^2,m_1,m_2)&=&{1\over6}a(m_1)+{1\over6}a(m_2)
+{1\over3}m^2_2~b_0(q^2,m_1,m_2)\nonumber\\
&-&{(q^2-m_1^2+m^2_2)^2\over12q^2}b_0(q^2,m_1,m_2)\\
&-&{m_1^2-m^2_2\over12q^2}\left[a(m_1)-a(m_2)\right]
+{1\over6}(m^2_1+m^2_2-{1\over3}q^2)\nonumber
\label{eqn:defb22}
\end{eqnarray}
Here ~$a$ ~and ~$b_0$ ~are the ``renormalized in $\overline{MS}$ scheme''
functions:
\begin{eqnarray}
a(m) = m^2(-1 + \log{m^2\over M_Z^2})
\end{eqnarray}
\begin{eqnarray}
b_0(q^2,m_1,m_2)
= -\int_0^1dx \log{x(x-1)q^2 + (1-x)m_1^2 + x m^2_2\over M_Z^2}
\label{eqn:defb0}
\end{eqnarray}

In terms of the mixing matrices ~$Z_{\pm}$ ~and ~$Z_N$ ~the couplings
needed in  eqs.(60,64) take the form \cite{ROS}
\footnote{Factors ~$c^{ji}_{L,N}$ ~and ~$c^{ji}_{R,N}$ ~given in \cite{ROS}
are not properly symmetrized.}:
(the fermion ~$i$ ~is incoming and ~$j$ ~is outgoing)
\begin{eqnarray}
c^{ji}_{L,C/N} &=& Z^{1j*}_+Z^{2i}_N - {1\over\sqrt2}Z^{2j*}_+Z^{4i}_N
\nonumber\\
c^{ji}_{R,C/N} &=& Z^{1j}_-Z^{2i*}_N + {1\over\sqrt2}Z^{2j}_+Z^{3i*}_N
\end{eqnarray}
\begin{eqnarray}
c^{ji}_{L,C} &=& Z^{1j*}_+Z^{1i}_+ + (\hat c^2- \hat s^2)\delta^{ji}
\nonumber\\
c^{ji}_{R,C} &=& Z^{1j}_-Z^{1i*}_- + (\hat c^2- \hat s^2)\delta^{ji}
\end{eqnarray}
and
\begin{eqnarray}
c^{ji}_{L,N} &=& Z^{4j*}_NZ^{4i}_N - Z^{3j}_NZ^{3i*}_N
\nonumber\\
c^{ji}_{R,N} &=& Z^{3j}_NZ^{3i*}_N - Z^{4j*}_NZ^{4i}_N
\end{eqnarray}

We end this Appendix with two comments. First, as long as neutralinos
appear only in closed loops it is not necessary to do anything with
negative eigenvalues of their mass matrix (53). Using ~
$(\not\! p + m)/(p^2- m^2)$ ~for ~$m<0$ ~as a propagator gives the same
physical results as other more complicated prescriptions which require
modification of couplings (see eg. \cite{HHG}). It is possible, however,
to choose the ~$Z_N$ ~matrix such that all mass eigenvalues are positive.

Second, the transition from Weyl spinors ~$\lambda^{0,\pm}_i$ ~to four
component fields ~$\psi_{N_i}$  ~and ~$\psi_{C^+_i}$ ~is not unique. One
can work for example with
\begin{eqnarray}
\psi_{C^-_i}= \left(\matrix{\lambda^-_i\cr\overline\lambda^+_i}\right)
\end{eqnarray}
instead of ~$\psi_{C^+_i}$. ~This freedom in building four component
fields from two component ones is in fact the basis of the recently
developped technique \cite{DEN} for handling Majorana particles which
often produce ``clashing arrows'' in Feynman diagrams. However realizing
that (at the level of lagrangian) e.g.
\begin{eqnarray}
a^{ji} \overline\psi_{C^+_j} (1-\gamma^5)\psi_{N_i} \phi^+=
a^{ji} \overline\psi_{N_i} (1-\gamma^5)\psi_{C^-_j} \phi^+
\end{eqnarray}
and that in different vertices of a given Feynman diagram different forms of
the same interaction can be used allow avoiding essentially all
problems with Majorana particles.

\newpage
\noindent {\bf FIGURE CAPTIONS}
\vskip 0.5cm

\noindent {\bf Figure 1.}

\noindent Comparison  of ~$\sin^2\theta(M_Z)_{FULL}$ ~
with ~$\sin^2\theta(M_Z)_{LLT}$ ~in the MSSM
for sparticle spectra corresponding
to the general scan described in the text in four different cases:\\
a) $m_t = 160$ GeV ~$\tan\beta = 2$,\\
b) $m_t = 160$ GeV ~$\tan\beta = 50$,\\
c) $m_t = 180$ GeV ~$\tan\beta = 2$,\\
d) $m_t = 180$ GeV ~$\tan\beta = 50$.\\
\noindent
Cirles denote spectra with all SUSY particles heavier than ~
500 GeV  ~and  ~$50 < M_{A^0} < 500$ GeV. ~
Stars correspond to spectra with ~
$M_{A^0} \leq 250$ GeV, ~$m_{C_i}, ~m_{N_j} \leq 250$ GeV ~
and heavy sfermions, ~$\geq500$ GeV. ~
Squares correspond to spectra with all sparticles light, $<250$ GeV
and ~$50 < M_{A^0} < 250$ GeV.
\vskip 0.3cm

\noindent {\bf Figure 2.}

\noindent a) ~$\sin^2\theta(M_Z)$ ~and b) ~$\alpha(M_Z)_{EM}$ ~in the MSSM
as a function of the (common for all the three generations) sneutrino
mass. All other SUSY particles (including the ~$A^0$~ Higgs boson)
have masses 2 TeV, with no Left -- Right mixing.
Solid lines correspond to the full calculation,
dashed ones to the LLT approximation, dot--dashed lines show our
approximation (eqs.(\ref{eqn:damy},\ref{eqn:dsmy}) combined with
the fit (\ref{eqn:fits})). Two dotted lines show the results of
application of the Faraggi Grinstein formulae (see \cite{FAGRI},
eqs. (5.10)) to the SM couplings
obtained from our fit (\ref{eqn:fits}).
\vskip 0.3cm

\noindent {\bf Figure 3.}

\noindent a) ~$\sin^2\theta(M_Z)$ ~and b) ~$\alpha(M_Z)_{EM}$ ~in the MSSM
as a function of the (purely) left handed sbottom mass. All other SUSY
particles (including the ~$A^0$~ Higgs boson) have masses ~2 TeV, ~with no
Left -- Right mixing. Solid lines correspond to the full calculation,
dashed ones to the LLT approximation, dot--dashed lines show our
approximation (eqs.(\ref{eqn:damy},\ref{eqn:dsmy}) combined with
the fit (\ref{eqn:fits})). Two dotted lines show the results of
application of the Faraggi Grinstein formulae (see \cite{FAGRI},
eqs. (5.10)) to the SM couplings obtained from our fit (\ref{eqn:fits}).
\vskip 0.3cm

\noindent {\bf Figure 4.}

\noindent a) ~$\sin^2\theta(M_Z)$ ~and b) ~$\alpha(M_Z)_{EM}$ ~in the MSSM
as a function of the pseudoscalar mass ~$M_{A^0}$. ~mass. All SUSY particles
have masses 2 TeV,  with no Left -- Right mixing. Solid lines correspond to
the full calculation, dashed ones to the LLT approximation, dot--dashed lines
show our approximation (eqs.(\ref{eqn:damy},\ref{eqn:dsmy}) combined with the
fit (\ref{eqn:fits})). Three dotted lines show the results of application of
the Faraggi Grinstein formulae (see \cite{FAGRI}, eqs. (5.10)) to the SM
couplings obtained from our fit (\ref{eqn:fits}).
\vskip 0.3cm

\newpage
\noindent {\bf Figure 5.}

\noindent
Parameter ~$\mu$ ~as a function of ~$T_{SUSY}$ ~for spectra of ref.
\cite{OLPOK,COPW} obtained for different values of ~$m_t$ ~and ~$\tan\beta$ ~
and subject to the conditions: ~i) Squark masses below 2 TeV, ~ii)
Sparticle masses above the experimental limits.
\vskip 0.3cm

\noindent {\bf Figure 6.}

\noindent
Values of ~$\hat\alpha_3$ ~and ~$\hat s^2$ ~(vertical and horizontal axis
respectively)  calculated with the use of sparticle spectra characterized
in Fig.5  and for the same ~$m_t$ ~and ~$\tan\beta$ ~values. Stars (diamonds)
denote LLT (full) calculation of the MSSM couplings. The upper (lower) group
of points in Figs. a), c) and d) corresponds to  ~$\hat\alpha_3^{SM}=0.125
(0.115)$. ~The three groups of points in Fig. b) correspond to ~
$\hat\alpha_3^{SM}=0.13,~0.125,~0.115$~.Contours in the plot denote bounds
imposed by SUSY SU(5) unification conditions with ~$\hat\alpha_{em}$ ~(in the
MSSM) allowed to vary in the range: ~1/129.0 -- 1/131.8: ~Solid lines - no GUT
scale thresholds; dashed (dash-dotted) lines --  $M_{GUT}^{max}/M_{GUT}^{min}=
30$ (1000) ~(for details see text). The three dotted lines correspond to fixed
values of ~$M_{H_3}= 1 \times 10^{17}$ ~(uppermost), ~$2\times 10^{16}, ~
5\times 10^{15}$ ~(for ~$\hat\alpha_{EM}=1/131.8$). ~The region below
the long--dashed lines correspond to violation of perturbativity
condition ~$Y_t^2(M_{GUT})/{4 \pi} <1$.
\vskip 0.3cm

\noindent {\bf Figure 7.}

\noindent
Values of ~$\hat\alpha_3^{SM}$ ~ predicted by SUSY unification without GUT
scale thresholds as a function of ~$T_{SUSY}$ ~calculated with the use of
sparticle spectra characterized in Fig.5  and for the same ~$m_t$ ~and ~
$\tan \beta$ ~values. Stars (diamonds) denote LLT (full) calculation of the
MSSM couplings. The region below the long-dashed line corresponds to violation
of perturbativity condition ~($Y_t^2(M_{GUT})/{4 \pi} <1)$.
\vskip 0.3cm

\noindent {\bf Figure 8.}

\noindent
$M_{H_3}$ ~as a function
of ~$T_{SUSY}$ ~for ~$\hat\alpha_3^{SM}(M_Z)=0.125$ ~
and sparticle spectra characterized
in Fig.5. Stars (diamonds) denote LLT
(full) calculation of the MSSM couplings.

\end{document}